\documentclass[3p,times]{elsarticle}

\usepackage{ecrc}

\volume{00}

\firstpage{1}

\journalname{Nuclear Physics A}

\runauth{}

\jid{nupha}

\usepackage{amssymb}

\usepackage[figuresright]{rotating}

\begin{document}

\begin{frontmatter}



\dochead{}

\title{The Rise and Fall of the Ridge}


\author{Paul Sorensen}

\address{Brookhaven National Laboratory, Upton, NY 11973}

\begin{abstract}
  Recent data from heavy ion collisions at RHIC show unexpectedly
  large near-angle correlations that broaden longitudinally with
  centrality. The amplitude of this ridge-like correlation rises
  rapidly with centrality, reaches a maximum, and then falls in the
  most central collisions. In this talk we explain how this behavior
  can be easily understood in a picture where final momentum-space
  correlations are driven by initial coordinate space density
  fluctuations. We propose $v_n^2/\varepsilon_{n,part}^{2}$ as a
  useful way to study these effects and explain what it tells us about
  the collision dynamics.
\end{abstract}


\end{frontmatter}

%
\textbf{Introduction}: The motivation for the construction of the
Relativistic Heavy Ion Collider at Brookhaven National Laboratory was
to collide heavy nuclei in order to form a state of matter called the
Quark Gluon Plasma (QGP) \cite{Reisdorf:1997fx}. These collisions
deposit many TeV into a region the size of a nucleus. The matter left
behind in that region is so hot and dense that hadronic matter
undergoes a phase transition into a form of matter where quarks and
gluons are the relevant degrees of freedom, not
hadrons~\cite{eos}. This is the state of matter that existed when the
universe was less than a microsecond old and still very hot.

Correlations and fluctuations are an invaluable tool for probing the
dynamics of heavy-ion collisions. Data from the experiments at RHIC
reveal interesting features in the two-particle correlation landscape
\cite{onset,ridgedata}. Specifically, it has been found that
correlation structures exist that are unique to Nucleus-Nucleus
collisions. While two-particle correlations in p+p and d+Au collisions
show a peak narrow in azimuth and rapidity, the near-side peak in
Au+Au collisions broadens substantially in the longitudinal direction
and narrows in azimuth. An analysis of the width of the peak for
particles of all $p_T$ finds the correlation extends across nearly 2
units of pseudo-rapidity ($\Delta\eta=2$)~\cite{ridgedata}. When triggering on
higher momentum particles ($p_T>2$~GeV/c for example), the correlation
extends beyond the acceptance of the STAR detector ($\Delta\eta<2$)
and perhaps as far as $\Delta\eta=4$ as indicated by
PHOBOS data~\cite{ridgedata}.

STAR has found that the amplitude of this correlation shows a rather
rapid rise with collision centrality~\cite{onset} before reaching a
maximum and falling off in the most central bins. This drop in the
most central bins shows up for both $\sqrt{s_{_{NN}}}=$ 200 and 62.4
GeV but is often overlooked. In this talk we present a geometric
explanation for the centrality dependence of the ridge
amplitude. We'll use the centrality dependence of
$v_2/\varepsilon_{2}$, $dN/dy$, and the third harmonic participant
eccentricity $\varepsilon_{3,part}^2$~\cite{v3} to predict the
amplitude of the near-side ridge correlation ($A_1$). Given the
apparent relevance of initial-state density fluctuations to the final,
momentum-space correlations, we advocate the transfer function
$v_n^2/\varepsilon_{n,part}^2$ as a valuable observable for studying
the length scales in heavy-ion collisions. We extract the transfer
functions from intermediate $p_T$ di-hadron correlation data where
evidence has been presented for conical emission~\cite{dihadrons}.


\textbf{Three Premises}: It has been shown by the STAR collaboration
that the second harmonic component of the near-side ridge in the two
particle correlations can account for nearly all of the difference
between the two and four particle cumulant $v_2$~\cite{glurad}. It was
argued that $v_2$ fluctuations must therefore be tiny: many times
smaller than the fluctuations predicted from eccentricity
models~\cite{glauber}. In that case, a major revision of our
understanding of heavy-ion collisions would be required. We've argued
previously, however, that eccentricity fluctuations can give rise to
$v_n$ fluctuations for more harmonics than just $n=2$, and that those
fluctuations could therefore be the source of the near-side
ridge~\cite{sorensen1} (especially if the $v_n$ fluctuations depend on
pseudo-rapidity difference $\Delta\eta$ \textit{i.e.} $\langle
v_n(\eta)v_n(\eta+\Delta\eta)\rangle \equiv f(\Delta\eta)$). This idea
has been born out by calculations from several
groups~\cite{mclerran,v3,brazil,hannah}. To test this
conjecture, we will attempt to explain the centrality dependence of the
near-side ridge amplitude $A_1$ from eccentricity fluctuations. we
start with three simple premises:
\begin{itemize}
\item the expansion of the fireball created in
  heavy-ion collisions converts anisotropies from coordinate-space into
  momentum-space,
\item the conversion efficiency increases with density,
\item and the relevant expansion plane is the participant plane.
\end{itemize}
The participant plane can be defined for any harmonic number and a
system with a lumpy initial energy density will give rise to finite
participant eccentricity at several harmonics~\cite{v3,teaneyv3}. This
becomes conceptually clear when eccentricity
is recast in terms of a harmonic decomposition of the azimuthal
dependence of the initial density. Our calculation of the centrality
dependence of $A_1$ will depend on the higher harmonic terms in an
eccentricity model.

\begin{figure}[htb]
 \centering
 \resizebox{0.75\textwidth}{!}{\includegraphics{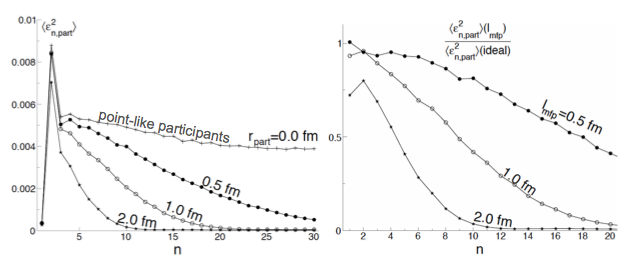}}
 \caption[]{ Left panel: $\langle\varepsilon_{n,part}^2\rangle$ for
   0-5\% central Au+Au collisions from a Glauber Monte Carlo where
   participants are either treated as point-like or they are smeared
   over a region of size $r_{part}$. We include the oblate shape of Au
   nuclei in our calculation. Right panel: the ratio of the curves on
   the left to the point-like curve.}
\label{f1}
\end{figure}

\textbf{Higher Harmonics, Even the Odd}: Fig.~\ref{f1} (left) shows
the $n^{th}$-harmonic participant eccentricity
$\langle\varepsilon_{n,part}^2\rangle$ as defined in Ref.~\cite{v3}
for central Au+Au collisions from a Monte-Carlo Glauber
model. Typically the participant eccentricity is calculated based on
the positions of point-like participants (that is the participant is
said to exist at a precise $x$ and $y$ position). Those results are
labeled $r_{part}=0.0$ fm. One can also calculate the eccentricity
from a more physically realistic model with participants smeared over
some region. This is done by treating each participant as many points
distributed within a disk of some finite radius $r_{part}$. Increasing
$r_{part}$ washes out the higher
$\langle\varepsilon_{n,part}^2\rangle$ terms. The right panel of
Fig.~\ref{f1} shows the ratio of
$\langle\varepsilon_{n,part}^2\rangle$ for a given $r_{part}$ value
divided by $\langle\varepsilon_{n,part}^2\rangle$ for
$r_{part}=0$. The curves are labeled $l_{mfp}$ instead of $r_{part}$
and the ratio $l_{mfp}/$ideal for reasons explained below.

\begin{figure}[htb]
  \centering
  \resizebox{0.65\textwidth}{!}{\includegraphics{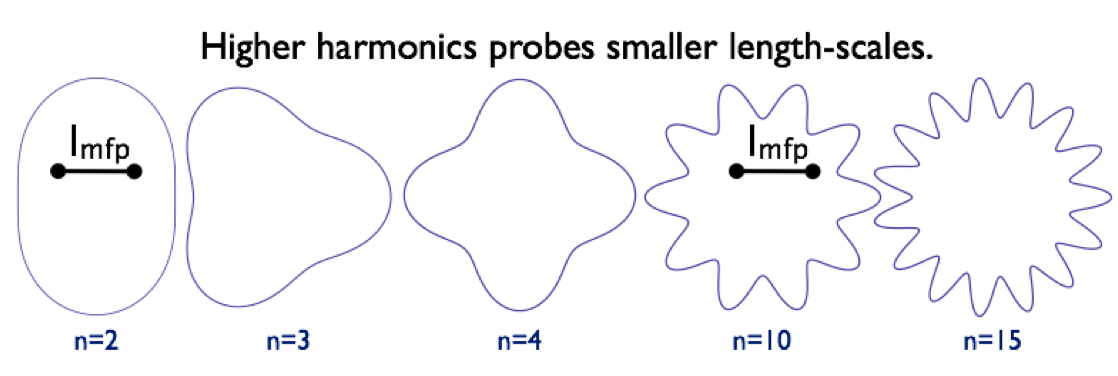}}
  \caption[]{ Schematic illustration of the interplay between a length
    scale like the mean-free-path and the higher harmonics. }
\label{f2}
\end{figure}

In this calculation we introduced the length scale $r_{part}$
causing the higher terms in $\langle\varepsilon_{n,part}^2\rangle$ to
be washed out. That effect is more general though and we believe it is
important for understanding correlations and $v_n$ fluctuations. One
can also consider what happens when particles free-stream for some
amount of time $\tau_{fs}$ before they interact; that will also
introduce a length scale $c\tau_{fs}$ which leads to a similar
reduction of higher terms~\cite{hannah}. One can also consider the
effect of a mean-free-path ($l_{mfp}$) on the ability of the fireball
to convert higher $\langle\varepsilon_{n,part}^2\rangle$ terms into
$v_n^2$~\cite{sound,ollialver}. If a particle on average travels for a distance
$l_{mfp}$ between interactions, it is clear that higher
$\langle\varepsilon_{n,part}^2\rangle$ terms will not become manifest
in $v_n^2$. Fig.~\ref{f2} shows a schematic illustration of this
idea. If our probe has a $l_{mfp}$ in the fireball, then we will
be blind to features smaller than $l_{mfp}$. All these
effects will act to wash out the higher harmonics so we expect
$v_n^2$ to drop with $n$. Since $v_n^2$ is related to $dN/d\Delta\phi$
by a Fourier transform, and since a Fourier transform of a Gaussian is
a Gaussian, all we need to reproduce the near-side Gaussian ridge in
two particle correlations, is for $v_n^2$ to drop with $n$ with an
approximately Gaussian shape. The right panel of Fig.~\ref{f1} shows
that this is a reasonable expectation from viscous effects.

\begin{figure}[htb]
  \centering
  \resizebox{0.5\textwidth}{!}{\includegraphics{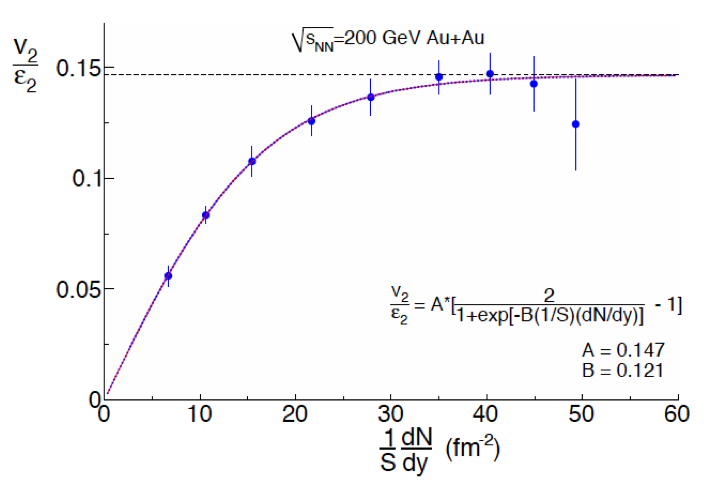}}
  \caption[]{ The ratio of $v_2\{4\}$ over $\varepsilon_{std}$ from a
    CGC model vs $(1/S)dN/dy$. }
\label{f3}
\end{figure}

The ridge amplitude $A_1$ is found by measuring
$\Delta\rho/\sqrt{\rho_{ref}}$ (the pair density $\rho$ minus the
reference pair density $\rho_{ref}$ scaled by $\sqrt{\rho_{ref}}$) vs
$\Delta\phi$ and $\Delta\eta$~\cite{daugherity}. A fit function is devised
to describe the correlation. The fit function has a $\Delta\phi$
independent term, a $\cos(\Delta\phi)$ and $\cos(2\Delta\phi)$ term, and a
near-side 2-D Gaussian. The fit function describes the data well~\cite{daugherity}. Here
we work with the conjecture that the 2-D Gaussian is a manifestation
of $\langle\varepsilon_{n,part}^2\rangle$. Based on this conjecture,
we can try to calculate the centrality dependence of $A_1$. Our result
for $A_1$ will be related to $v_n^2$ so we need to know the conversion
efficiency $c$ of $\langle\varepsilon_{n,part}^2\rangle$ into
$v_n^2$. The conversion efficiency will depend on particle density. In
Fig.~\ref{f3} we show $v_2/\varepsilon$ vs density $(1/S)dN/dy$ from
the STAR Collaboration~\cite{v2papers} with a fit function to parameterize the
data. The figure shows $v_2\{4\}$ divided by the eccentricity
calculated with respect to the reaction plane from a CGC model~\cite{fklncgc}. We
will take this to estimate our conversion efficiency with the
understanding that the different results obtained from CGC and Glauber
models gives rise to at least a 30\% uncertainty in the correct
conversion efficiency.

Since a $\cos(\Delta\phi)$ and $\cos(2\Delta\phi)$ term have already
been subtracted from $\Delta\rho/\sqrt{\rho_{ref}}$ in order to obtain
$A_1$, our estimate of $A_1$ can be made simpler if we predict the
$n=3$ component of the near-side ridge and scale that up to get the
full amplitude. The $n=3$ component is found from
\begin{equation}
\frac{1}{2\pi}\int_{-\pi}^{\pi}\frac{\Delta\rho}{\sqrt{\rho_{ref}}}\cos(3\Delta\phi)d\Delta\phi
= 0.039A_1 \,.
\end{equation}
The azimuthal width $\sigma$ of the near-side Gaussian is weakly
dependent on centrality so we use a typical value of
$\sigma=0.65$. The factor of 0.039 will be used to relate our
prediction for $\langle\varepsilon_{n,part}^2\rangle$ to the amplitude
$A_1$. Since $\Delta\rho/\sqrt{\rho_{ref}}$ is a per-particle measure
instead of a per-pair measure, we need to include the particle density
$\rho_0=\frac{1}{4\pi}\frac{dN}{dy}$. Putting
$\langle\varepsilon_{3,part}^2\rangle$ together with the conversion
efficiency $c$, particle density $\rho_0$, and the factor of 0.039 to
map from the $n=3$ component to a Gaussian amplitude, we find
\begin{equation}
A_1 \approx \rho_0c\varepsilon_{3,part}^2/0.039 \, .
\end{equation}
We take $\rho_0$ from data, $c$ from Fig.~\ref{f3}, and
$\langle\varepsilon_{3,part}^2\rangle$ from our Monte-Carlo Glauber
model.

\begin{figure}[htb]
  \centering
  \resizebox{0.75\textwidth}{!}{\includegraphics{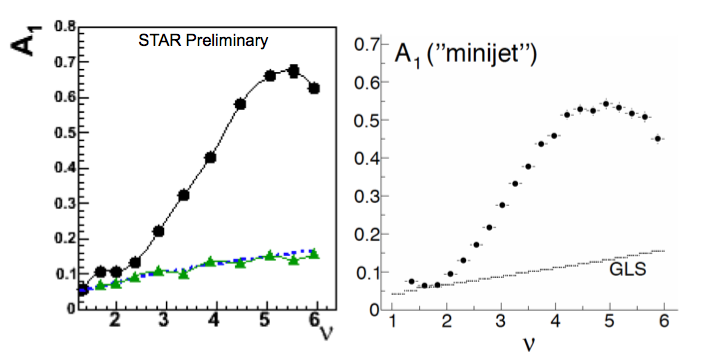}}
  \caption[]{ Left panel: measured near-side ridge amplitude $A_1$ vs
    centrality measure $\nu = 2N_{bin}/N_{part}$. Right panel:
    contribution to the ridge from initial state geometry
    fluctuations. }
\label{f4}
\end{figure}

\textbf{The Rise and Fall of the Ridge}: The right panel of
Fig.~\ref{f4} shows our estimate of the ridge amplitude $A_1$ based on
$\varepsilon_{n,part}^2$ vs centrality parameter
$\nu=2N_{bin}/N_{part}$. The left panel shows the preliminary STAR
data. Our estimate for the amplitude is only approximate since we've
applied the conversion efficiency for $n=2$ to $n=3$ and as discussed
before, we expect the efficiency to drop with $n$. This will lead us
to overestimate the contribution from
$\langle\varepsilon_{n,part}^2\rangle$ to the ridge. On the other
hand, we used a CGC based eccentricity model to extract the conversion
efficiency but a Glauber model for
$\langle\varepsilon_{3,part}^2\rangle$. This should cause us to
under-estimate the ridge contribution; to some extent canceling the
previous overestimate. We find that our estimate of $A_1$ agrees
quite well with data. The centrality dependence is particularly
interesting: our $A_1$, like the data, starts at a small value and
rises much faster than expectations from a Glauber Linear
Superposition model (GLM) which assumes that correlations grow as
$N_{bin}/N_{part}$. The rise continues until $A_1$ reaches a maximum
near $\nu=5$, then $A_1$ falls again. This rise and fall is also seen
in the preliminary 62.4 GeV data~\cite{daugherity} and has no natural
explanation in any of the other proposed scenarios for the ridge
formation. In our picture, the rise and fall is related simply to the
geometry and it's fluctuations.
$\langle\varepsilon_{3,part}^2\rangle$ falls with $N_{part}$ since the
larger $N_{part}$ leads to smaller fluctuations. But
$N_{part}\langle\varepsilon_{3,part}^2\rangle$ first rises then
falls. This rise and fall is due the asymmetry of the overlap region
which shows up even for $\langle\varepsilon_{3,part}^2\rangle$. Since
both $c$ and $\rho_0$ are increasing with centrality, the product of
$\rho_0c\langle\varepsilon_{3,part}^2\rangle$ rises until very
central collisions and then falls as shown in the figure. The
observation that the rise and fall shows up in the near-side ridge
amplitude suggests that the near-side ridge is likely dominated by initial
geometry fluctuations.

\textbf{Transfer Functions}: Our explanation for the centrality
dependence of $A_1$ is based on the three simple premises listed
earlier and provides a natural explanation for the rise and fall of
the ridge. Initial estimates of the amplitude agree to within our
uncertainties. This suggests that geometry fluctuations in the initial
overlap region are converted into momentum space giving rise to the
near-side ridge structure. We discussed that we expect the conversion
efficiency to drop with $n$ since effects like initial-state
free-streaming and mean-free-path will wash out the higher harmonic
terms. Measuring the conversion efficiency
$c_n=\frac{v_n\{2\}^2}{\varepsilon_{n,part}\{2\}^2}$ as a function of
$n$, centrality, and particle kinematics will provide information on
those effects. It will be particularly interesting to measure $c_n$ as
a function of $\Delta\eta$ to understand how de-coherence affects
manifest in the longitudinal direction.

\begin{figure}[htb]
  \centering
  \resizebox{0.85\textwidth}{!}{\includegraphics{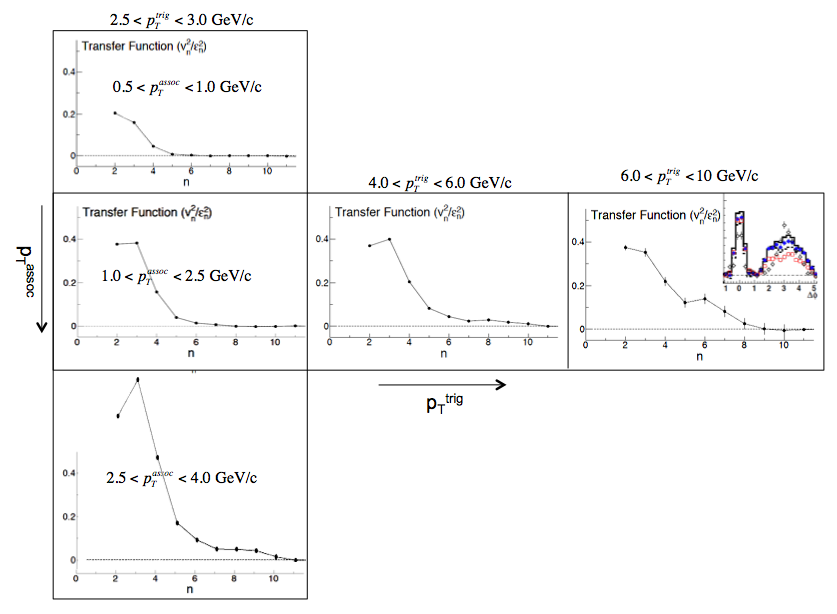}}
  \caption[]{ The transfer function
    $v_n\{2\}^2/\varepsilon_{n,part}^2$ for intermediate $p_T$
    di-hadron correlations from 20-60\% central 200 GeV Au+Au
    collisions~\cite{dihadrons}.}
\label{f5}
\end{figure}

In Fig.~\ref{f5} we show the conversion efficiency
$v_n\{2\}^2/\varepsilon_{n,part}^2$ for intermediate $p_T$ di-hadron
correlations from STAR~\cite{dihadrons}. The panels show various
combinations of trigger particle and associated particle $p_T$
selections. Intermediate $p_T$ di-hadron data rather than showing an
away-side Gaussian, show two peaks shifted to either side of
$\Delta\phi=\pi$. It has been suggested that this is evidence of
conical emission on the away-side of the higher $p_T$ trigger
particle~\cite{awayside,mach}. The previously ignored effects of
$v_3^2$ which were, however, could explain some or all of these novel
structures~\cite{sorensen1}. To investigate these structures, we look
at the $p_T$ dependence of $c_{n}(p_T^{trig},p_T^{assoc})$. By
plotting $v_n\{2\}^2/\varepsilon_{n,part}^2$ from these correlations,
we hope to see what portion of those correlations can be explained by
geometric effects. For relatively lower momentum cuts, we find that
$c_{n}$(2.5 GeV, 0.5 GeV) behaves much as we expect from mean-free
path effects for example; the higher terms drop off monotonically
similar to Fig.~\ref{f1} (right). But for higher $p_T^{assoc}$ cuts
$c_n$ shows a pronounced peak at $n=3$. See $c_n$(2.5 GeV, 2.5 GeV)
for example. This feature disappears again however when a large enough
$p_T^{trig}$ cut is applied. For $p_T^{trig}>6$ GeV for example, we
see only effects due to the presence of correlations from jets. It
remains interesting to speculate about the possible source of that
local maximum at $n=3$ for intermediate $p_T$ di-hadron correlations
(whether it's due to a suppression of the lower harmonic super-horizon
modes~\cite{Mishra:2007tw}, mach-cones~\cite{mach} or some other
acoustic effects~\cite{andrade}). It seems to be larger than trivially
expected from $\varepsilon_{n,part}^2$. We note however that a
complete investigation of the systematic errors on our Glauber Model
has not been carried out. We include the oblate shape measured for the
Au nucleus. That oblateness leads to a large enhancement of the $n=2$
component of $\varepsilon_{n,part}^2$. Reducing the oblateness of the
Au nucleus could therefore reduce or eliminate the prominence of the
$n=3$ term in $c_n$. This remains for further investigation.

\textbf{Conclusions}: We presented the participant eccentricity vs
harmonic when the participants are treated as point-like or smeared
over a radius $r_{part}$. The larger values of $r_{part}$ wash out the
higher harmonic eccentricities. We argued that, similarly, a large
mean-free-path should wash out higher harmonics of $v_n$. Such an
effect could lead to a Gaussian peak in two particle correlations at
$\Delta\phi=0$.  We've calculated the contribution to the near-side
Gaussian peak that we expect from initial density fluctuations. We
based our calculation on three premises 1) that the expansion of the
fireball created in heavy-ion collisions converts anisotropies from
coordinate-space into momentum-space, 2) the conversion efficiency
depends on particle density, and 3) the relevant expansion plane is
the participant plane. Following these premises, we find that the
near-side peak from density fluctuations should rise rapidly, reach a
maximum just before the most central events, then fall. Our estimate
of the magnitude is in agreement within our uncertainties with the
available data and the shape matches that seen in the data. This is
the only calculation to correctly describe the rise and fall of the
ridge amplitude. We conclude therefore that density fluctuations are
likely the dominant source for the low $p_T$ ridge-like correlations.
We have also shown the ratio of the final momentum space correlations
$v_n\{2\}^2$ to the initial coordinate-space eccentricities
$\varepsilon_{n,part}^2$ for intermediate $p_T$ di-hadrons. We find
that for intermediate $p_T$ correlations the $n=3$ term is larger than
$n=2$. This suggests that even after taking into account initial
density fluctuations, some interesting signal may exist in the
intermediate $p_T$ di-hadron data.

\end{document}